\documentclass[fleqn,twoside]{article}
\usepackage{espcrc2}

\usepackage{xspace}
\usepackage{graphicx}
\usepackage{dcolumn}
\usepackage{amsmath}
\usepackage{epsfig}

\input babarsym.tex

\newcommand{\roots}        {\ensuremath{\sqrt{s}}\xspace}

\newcommand{\eetautau}   {\ensuremath{e^+e^-\to\tautau}\xspace}
\newcommand{\eemm}       {\ensuremath{e^+e^-\to\mumu}\xspace}
\newcommand{\eeqq}       {\ensuremath{e^+e^-\to\qqbar}\xspace}

\newcommand{\tautomu}  {\ensuremath{\tau \to \mu \nunub}\xspace}
\newcommand{\tautoel}  {\ensuremath{\tau \to e \nunub}\xspace}

\newcommand{\tautohnpiz}  {\ensuremath{\tau \to h (\ge 1) \pi^0 \nu}\xspace}

\newcommand{\taumg}      {\ensuremath{\mtau^{\pm} \to \mmu^{\pm} \g}\xspace}
\newcommand{\muelg}      {\ensuremath{\mmu^{\pm} \to \electron^{\pm} \g}\xspace}

\newcommand{\BRtaumg}    {\ensuremath{\BR(\taumg)}\xspace}
\newcommand{\BRmuelg}    {\ensuremath{\BR(\muelg)}\xspace}

\newcommand{\ntaupair}         {\ensuremath{1.97\times 10^8}\xspace}

\newcommand{\lumion}             {\ensuremath{205.4 \invfb}\xspace}
\newcommand{\lumioff}             {\ensuremath{16.1 \invfb}\xspace}

\newcommand{\sigeff}            {\ensuremath{(7.45\pm0.65)\%}\xspace}

\newcommand{\sigback}            {\ensuremath{6.2\pm 0.5}\xspace}
\newcommand{\EventsInBox}            {\ensuremath{4}\xspace}

\newcommand{\UpperLimitBayes}       {\BRtaumg$<14\times10^{-8}$ at 90\% CL\xspace}

\newcommand{\datatomcegsb}       {\ensuremath{1.048\pm0.055(stat)\pm0.023(norm)}\xspace}
\newcommand{\Emg}             {\ensuremath{E_{\mu\gamma}}\xspace}
\newcommand{\Mnu}             {\ensuremath{m_{\nu}^2}\xspace}

\newcommand{\logmisspt}             {\ensuremath{-\ln (2\times p^T_{miss}/\sqrt{s})\xspace}}

\newcommand{\trkhel}           {\ensuremath{\cos\theta_H}\xspace}

\def\tautomu   {\ensuremath{\tau^\pm \to \mu^\pm \nunub}\xspace}

\def\kk2f       {\mbox{\tt KK2f}\xspace}
\def\tauola     {\mbox{\tt Tauola}\xspace}

\newcommand{\gevccgevcc}{\ensuremath{{\mathrm{\,Ge\kern -0.1em V^2\!/}c^4}}\xspace}

\def\figurebox#1#2#3{%
    \def\arg{#3}%
    \ifx\arg\empty
    {\hfill\vbox{\hsize#2\hrule\hbox to #2{\vrule\hfill\vbox to #1{\hsize#2\vfill}\vrule}\hrule}\hfill}%
    \else
    {\hfill\epsfbox{#3}\hfill}%
    \fi}

\begin{document}

\pagestyle{empty}

\renewcommand{\thefootnote}{\fnsymbol{footnote}}

\begin{titlepage}
%%%%%%%%%%%%%%%
%% Substitute your Pub number, month and year
%% in the following
%%%%%%%%%%%%%%%

\begin{flushright}
{\small
SLAC--PUB--10872\\
BABAR-PROC-04/131\\
hep-ex/0412002\\
November 2004\\}
\end{flushright}

\vspace{.8cm}

%%%%% Title and Author Information:
%%
\begin{center}
{\large\bf
Search for Lepton Flavour Violation in the
 Decay $\tau^{\pm} \to \mu^{\pm} \gamma$}

\vspace{1cm}

J. M. Roney  \\
 Department of Physics and Astronomy,\\
 University of Victoria, \\
 P.O.Box 3055 STN CSC,\\ Victoria, B.C., V8W 3P6, CANADA\\
(representing the BABAR Collaboration)

\end{center}

\vfill
%%%%% If your paper has an abstract, put it here:
%%
\begin{center}
{\large\bf
Abstract }
\end{center}

\begin{quote}
A search for the lepton flavour violating decay $\tau^{\pm} \rightarrow \mu^{\pm} \gamma$
has been performed using 221.4{\ensuremath{\mbox{\,fb}^{-1}}}
 of data collected at an e$^+$e$^-$
centre-of-mass energy of 10.58~GeV with the BABAR detector
at the PEP-II storage ring.
The search has an efficiency of 7.45$\pm$0.65\% for an expected background
level of 6.2$\pm$0.5 events. In the final sample 4 candidate events
are selected. As there is no evidence for a  signal in this data, for this
preliminary result we set an upper limit of
${\cal B}(\tau^{\pm} \to \mu^{\pm} \gamma)< 9\times10^{-8}$ at 90\%~CL
using the method of Feldman and Cousins.
\end{quote}

\vfill

%%%%%%%%%%%%%%%
\begin{center}
To be published in the Proceedings of the \\ 8th International Workshop On Tau Lepton Physics\\
14-17 September 2004, Nara, Japan 
\vfill

\textit{Stanford Linear Accelerator Center, Stanford University,Stanford, CA  94309}\\
\vspace{2mm}
\rule[2mm]{12cm}{0.3mm} \\
Work supported by Department of Energy contract  DE--AC02--76SF00515.

\end{center}

\end{titlepage}
%\title{}
%\maketitle
\twocolumn

%%%%%%%%%%%%%%%%%%%%%%%%%%%%%%%%%%%%%%%%%%%%%%%%%%%
%%
%%   Sample draft of PRL in RevTEX format
%%   
%%%%%%%%%%%%%%%%%%%%%%%%%%%%%%%%%%%%%%%%%%%%%%%%%%%

%%
%% --------- Introduction ----------------
%%

 Decays violating lepton flavour number, if observed, would be
 among the most theoretically clean signatures of new physics.
 The decay \taumg is one such  process and is expected with potentially observable rates in 
 supersymmetric models\cite{Barbieri:1994pv,Hisano:1995nq,Hisano:1998fj},
 left-right supersymmetric models\cite{Babu:1999ge} and 
 supersymmetric string unified models\cite{King:1998nv}.  
 For some ranges of model parameters, decay rates as high as several parts per million 
 are expected for this decay\cite{Hisano:1998fj,King:1998nv}, 
 even in light of the current experimental limit on  the related \muelg decay\cite{Brooks:1999pu}. 
 The recently reported non-zero charge-parity asymmetry in b-s radiative penguin processes\cite{ichep04} 
 also suggests large \BRtaumg in grand unified theories with supersymmetry\cite{Hisano04}.
 On the other hand, the modest extensions to the standard model (SM) incorporating neutrino
 oscillations predict a  branching ratio
 many orders of magnitude below experimental accessibility ($\approx 10^{-40}$\cite{Pham:1998fq}).
 Therefore a discovery of this decay would require new physics and a non-observation of the decay
 would place restrictions on the parameters in the theories predicting large branching ratios.
 Currently the most stringent limit is \BRtaumg$<3.1\times10^{-7}$ at 90\% confidence level (CL)
 from the BELLE experiment\cite{Abe:2003sx} using 86.3\invfb of \epem annihilation data.

%%
%% --------- Data set and Detector ----------------
%%
The search for \taumg decays reported here uses data recorded by the \babar\ detector at the \pep2
 asymmetric-energy \epem storage ring operated at the Stanford Linear Accelerator Center.
 The data sample consists of an integrated luminosity of \L=\lumion\ recorded at a centre-of-mass energy (\roots) of
10.58\gev\ and  \lumioff\ recorded at  $\sqrt{s}=10.54\gev$ .
With a cross section for \eetautau at the luminosity-weighted 
$\sqrt{s}$ of $\sigma_{\tau\tau} = (0.89\pm0.02)$ nb\cite{kk},
this data sample contains \ntaupair \eetautau events.
 
The \babar\ detector is described in detail in Ref.\cite{detector}.
Charged particles are reconstructed as tracks with 
a 5-layer silicon vertex tracker and a 40-layer drift chamber (DCH)
inside a 1.5-T superconducting solenoidal magnet.
An electromagnetic calorimeter (EMC) consisting of 6580 CsI(Tl) 
crystals is used to identify electrons and photons. A ring-imaging
Cherenkov detector is used to identify charged hadrons.
The flux return (IFR) of the solenoid, instrumented with resistive 
plate chambers, is used to identify muons.

%%
%% --------- Selection ----------------
%%
The signature of the signal process is the presence of an isolated \mmu and \g 
having an invariant mass consistent with that of the \mtau (1.777\gevcc\cite{bes}),
the energy of the \mmu and \g (\Emg) equal to \roots/2 in the event centre-of-mass (CM) frame, 
and the characteristics of the other particles in the event consistent with a SM \mtau decay.
Such events are simulated with higher-order radiative corrections using \kk2f \cite{kk}
where one $\tau$ decays into \mmu\g with a flat phase space distribution\cite{flatphasespace}, while the other $\tau$ decays
according to measured rates\cite{PDG} simulated with \tauola\cite{tauola,photos}.
The detector response is simulated with \mbox{\tt GEANT4}~\cite{geant}. 
The simulated events for signal as well as SM background processes\cite{kk,tauola,photos,Lange:2001uf,Sjostrand:1995iq}
are then reconstructed in the same manner as data.
The dominant backgrounds are from SM \eemm($\gamma$)  and \eetautau($\gamma$) events.

Events with two or four well reconstructed tracks and zero net charge are selected.
The thrust, calculated with all observed charged and neutral particles, is required to lie 
between 0.9 and 0.975 to suppress \eeqq background with low thrust and \eemm and Bhabha backgrounds with
thrust close to unity.
In order to ensure the presence of at least one $\nu$ within the acceptance of the detector,
the lab-frame polar angle ($\theta_{miss}$) of the missing momentum of the event is required to lie 
within the geometrical acceptance of the detector ($-0.76<\cos \theta_{miss} <0.92$)
and the missing CM transverse momentum ($p^{T}_{miss}$) is required to be significantly  above zero:
$\logmisspt < 2 (4)$ for events with two (four) tracks.

The signal-side hemisphere, defined with respect to the thrust axis, is required to
contain one track with  CM momentum less than 4.5\gevc\  and
 at least one \g\ with a CM energy greater than 200\mev . 
The track must be identified  as a \mmu using DCH, EMC and IFR information 
and the \g\ candidate is the one which gives the mass of the \mmu\g system closest to the \mtau mass.
The resolution of the mass of \mmu\g is improved by using a kinematic fit with 
\Emg\ constrained to \roots/2 and by assigning 
the point of closest approach of the \mmu track from the \epem collision axis
to the origin of the \g candidate. This energy-constrained mass (\mec) and \DeltaE=\Emg$-$\roots/2 are
independent variables in the absence of initial state radiation (ISR).
The mean and width of the Gaussian core of the \mec and \DeltaE distributions for reconstructed 
signal events are: $\langle \mec \rangle$ = 1777\mevcc, $\sigma(\mec)$ = 9\mevcc, 
$\langle \DeltaE \rangle$ = -9\mev, $\sigma(\DeltaE)$ = 45\mev, 
%where the shift in \DeltaE is due to ISR.
where the shift in \DeltaE to low values comes from the tail induced by ISR.

Potential biases in the analysis are minimized by blinding
the data events within a $3\sigma$ ellipse centred around
$\langle \mec \rangle$ and $\langle \DeltaE \rangle$
% the expected peak of the two-dimensional distribution 
in the \mec-\DeltaE plane 
until all the optimization and systematic studies of the selection criteria 
have been completed.

In order to suppress \eemm($\gamma$) and \eetautau($\gamma$) events containing energetic
 final state radiation and radiation in 
\tautomu\ decays, an isolation criterion is imposed on the \mmu by requiring $|\trkhel| < 0.8$,
where $\theta_H$ is the angle between the \mmu momentum  in the $\tau$  rest frame
 and the $\tau$ momentum as measured in the lab frame.
Background contamination arising from \tautohnpiz decays with the hadronic track $(h)$ mis-identified as a \mmu
is reduced by requiring the sum of the CM energy of non-signal photon candidates
in the signal-side to be less than 200\mev.
If the reconstructed neutral particle identified as the signal photon has at least a
 1\% likelihood of 
arising from overlapping daughters in $\pi^0 \rightarrow \g\g$ decays, the event is removed
from consideration.

The tag-side hemisphere, which is expected to contain a SM \mtau decay, is required 
to have a total invariant mass  less than 1.6\gevcc 
and a CM momentum for each track less than 4.0\gevc 
to reduce background from \eeqq and \eemm processes, respectively.
The  $\qqbar$ background is further reduced by requiring the hemisphere to have
no more than six photon candidates.

%The tag-side hemisphere is also explicitly required to contain particles that are consistent
%with coming from semi-exclusive SM  $\tau$ decays.
 A tag-side hemisphere containing a single track is
 classified as e-tag,  $\mu$-tag or  h-tag 
if the total photon CM energy in the hemisphere is no more
 than 200\mev and the track is exclusively identified as an electron (e-tag), as a muon ($\mu$-tag)
 or as neither (h-tag). If the total photon CM energy in the hemisphere is more
 than 200\mev, then events are selected if the track is exclusively identified as an electron
 (e$\gamma$-tag) or as neither an electron nor as a muon (h$\gamma$-tag).
 These allow for the presence of radiation in \tautoel decays and for photons from 
 $\pi^0 \rightarrow \g\g$ in \tautohnpiz decays.
 If the tag-side contains three tracks, the event is classified as a 3h-tag.
 We explored other tag-side channels but the sensitivity of the search does not improve
 by including them.

For the hadronic tagging modes, the invariant mass squared (\Mnu) of the 
tag-side unobserved particle (assumed to be a \nut) can be well reconstructed 
assuming that the candidate \taumg decay fully reconstructs the direction of $\tau$ 
in the tag-side with the energy of the $\tau$ given by $\sqrt{s}/2$. 
$|\Mnu|$ is required to be less than 0.4\gevccgevcc for h-tag and 3h-tag events and
less then 0.8\gevccgevcc for h$\gamma$-tag events.

At this stage of the analysis 15\% of the \taumg signal events survive within 
a Grand Side Band (GSB) region defined as: $\mec \in [1.5, 2.1] \gevcc, \DeltaE \in [-1.0, 0.5] \gev$.
In the non-blinded parts of the GSB,  4489 events survive in the data
which agrees at the 5\% level with the Monte Carlo (MC) background expectation of 4709 events.
 80\% of MC are \eetautau
events, 82\% of which are \tautomu decays in the signal hemisphere.

To further suppress the backgrounds,
separate neural net (NN) based discriminators are employed
 for each of the six tag-side channels. Five observables are used as input to the NN:
the missing mass of the event,
% calculated as $\sqrt{s - \Pmiss^2}$,
the CM momentum of the highest momentum tag-side track, \trkhel, \logmisspt~and \Mnu.
Each NN is trained using data in the non-blinded part of the GSB to describe the background 
and $\mu\gamma$ MC in the full GSB region to describe the signal.
The NN input distributions of the data are in good agreement
 with MC backgrounds both in shape and absolute rates as are the distributions of the NN outputs.
The MC is then used to determine the cut values to be applied to the NN outputs by
optimizing on the expected 90\%CL upper limit\cite{Feldman:1997qc} 
for observing a signal inside a $2\sigma$ ellipse in \mec-\DeltaE plane centred around  $\langle \mec \rangle$ 
and $\langle \DeltaE \rangle$.

\begin{figure}[htb]
  \begin{center}
   \includegraphics[height=.4\textheight,width=.5\textwidth]{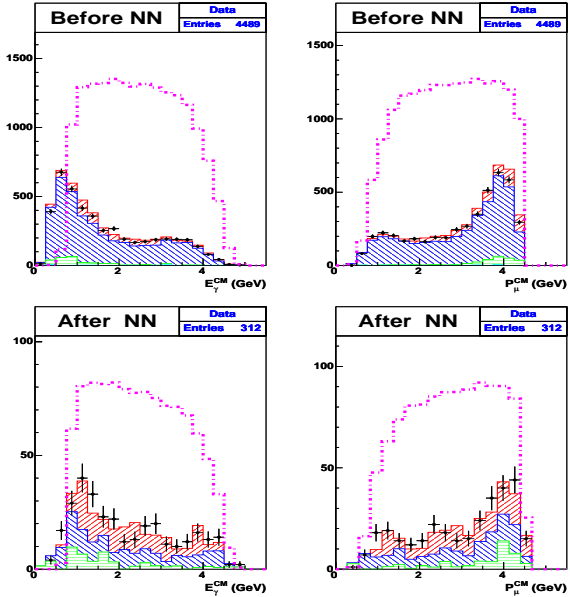}
    \caption[E$^{CM}_{\gamma}$ and P$^{CM}_{\mu}$ in GSB Regions before and after NN requirements applied]
          {E$^{CM}_{\gamma}$ and P$^{CM}_{\mu}$ in GSB Regions before and after the NN selection.
           Data are represented by points, background MC by hatched histograms (\eemm is on top, \eetautau below it
           and \eeqq on the bottom)  and signal MC by dash-lined histograms.}
    \label{fig:fig1}
  \end{center}
\end{figure}

\begin{figure}[htb]
  \begin{center}
   \includegraphics[height=.4\textheight,width=.5\textwidth]{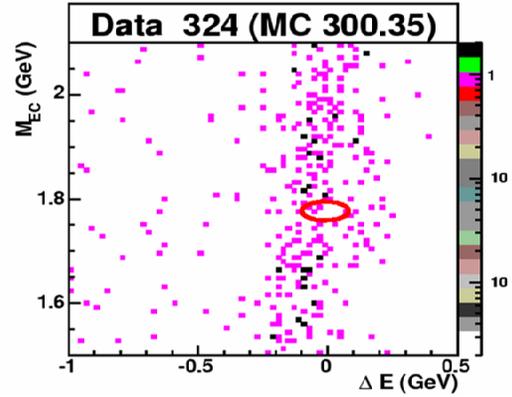}
    \caption[\mec {\em vs} \DeltaE in the GSB Regions after NN requirements applied]
          {\mec ~{\em vs} \DeltaE for data in all tag-side channels after applying the NN requirements. 
        The number of events in the data and in the MC samples are quoted. The ellipse depicts the $2\sigma$ signal region.}
    \label{fig:fig2}
  \end{center}
\end{figure}

The CM energy and momentum distributions of the \g and \mmu candidates
 before and after the NN selection has been applied
 are plotted in Figure~\ref{fig:fig1}.
 These distributions are also well described by the Monte Carlo simulation
 in both shape and rates. The two dimensional plot of  \mec {\em vs} \DeltaE for
 data is shown in Figure~\ref{fig:fig2}
 after the applying the NN selections.

\begin{figure*}[htb]
  \begin{center}
   \includegraphics[height=.4\textheight,width=.6\textwidth]{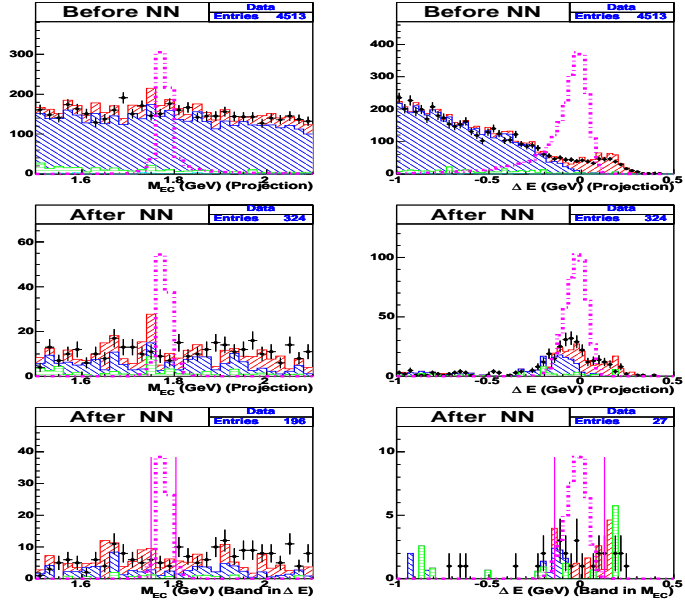}
    \caption[\mec and \DeltaE in the full GSB Regions before and after NN requirements applied]
          {\mec and \DeltaE in the full GSB Regions before and after the optimal NN requirement applied for
           all tag-side channels combined.
           Data are represented by points, background MC by hatched histograms,
           and signal MC by dash-lined histograms (with arbitary normalization).
           The selected number of events in the data is indicated in the upper right corner.
           The vertical lines indicate the 3$\sigma$ binded region which is removed for the background estimation.
In the lower left plot events lie within the band: $|\DeltaE - \langle\DeltaE\rangle| < 3\sigma(\DeltaE)$ 
and in the lower right plot events lie within the band:$|\mec - \langle\mec\rangle| < 3\sigma(\mec)$. }
    \label{fig:fig3}
  \end{center}
\end{figure*}

The distributions for \mec and \DeltaE before and after applying the  NN selection
 are shown in Figure~\ref{fig:fig3} where all channels have been combined.
 It is evident that there is good agreement between the data and MC in these
 observables.
 Because the missing mass is correlated with \DeltaE, after the NN selection is
 applied the remaining events tend to cluster towards low values of \DeltaE.
Also shown in Figure~\ref{fig:fig3} is the
 distribution of \mec  for events in
$|\DeltaE - \langle\DeltaE\rangle| < 3\sigma(\DeltaE)$
 and the
 distribution of \DeltaE  when \mec is restricted to
$|\mec - \langle\mec\rangle| < 3\sigma(\mec)$.

The number of background events is estimated after all criteria are applied except that on \mec.
The estimate uses the non-blinded region inside the band in \DeltaE :
$|\DeltaE - \langle\DeltaE\rangle| < 3\sigma(\DeltaE)$, which we refer to as the
 \mec sideband.
As is evident from  Figure~\ref{fig:fig3} these events are fairly
  uniformly spread in \mec and we estimate of the number of events
in the blinded signal region by scaling the number of events in the sideband
by the ratio of the area of the signal box to that of the sideband.

 For the MC in the 3$\sigma$ blinded region,
 the background interpolation predicts 11.4$\pm$0.9 events and 
 9.0 events are selected. This is to be compared with the data prediction of 14.0$\pm$1.0 events
 in this region.
 These 9.0 MC events correspond to 5.9 \eemm events,  2.5  \eetautau events and 0.7 uds events.
 The  \tautomu decays account for four of the five \eetautau background events observed in the MC.
 The expectations for the number of background events in each tag-side channel are in good agreement with
 the number selected in the MC. Alternative means for estimating the background
using the data yield consistent expected numbers of background events.

 The numbers of events in data and MC in the GSB for the different tag-side channels are
 presented in Table~\ref{tab:nnmubit7-2sig} both before and after the NN selection. Also included
 in the table are the numbers of events in the MC that are selected in the 2$\sigma$ signal region, the number of
 events in the signal region predicted from the MC  sideband and the number of events
 predicted from the data sideband.
 There is reasonable agreement between the number of events observed in the  data and
 those predicted by the MC for all tag-side channels both before and after the NN selection.

The relative systematic uncertainties on the trigger efficiency,
 tracking and photon reconstruction efficiencies, and particle identification 
are estimated to be 1.2\%, 1.3\%, 1.6\% and 1.2\%, respectively.
We evaluate an uncertainty on the efficiency arising from the NN selection
by  fixing each NN input variable to its average value one at a time, 
without changing the architecture of the NN, and re-calculating the efficiency. This has the
effect of removing each input variable completely from the NN selection procedure.
This results in 1.7\% relative variation 
of the signal efficiency. Adding these errors in quadrature yields a combined systematic
error estimate of 3.2\%.

%\begin{sidewaystable}[htbp]
\begin{table*}[htb]
\begin{center}
\caption[Optimized NN Selection for NN Tight $\mu$ PID selector]
{Tag-side dependence of the efficiency and no. events in the GSB, 3$\sigma$ blinded region and 2$\sigma$ signal
 region.}
{\scalebox{0.87}{\begin{tabular}{|c|c|c|c|c|c|c|c|c|c|c|c|} \hline
&& \multicolumn{2}{|c|}{MC Backgrounds} &\multicolumn{2}{|c|}{MC Backgrounds}           & \multicolumn{2}{|c|}{Data}&
\multicolumn{2}{|c|}{Data} & \multicolumn{2}{|c|}{Data} \\
&$\mu\gamma$& \multicolumn{2}{|c|}{Grand Side-Band}&\multicolumn{2}{|c|}{Signal Region}&
 \multicolumn{2}{|c|}{Grand Side-Band}& \multicolumn{2}{|c|}{3$\sigma$ blinded Region} & \multicolumn{2}{|c|}{Signal Region} \\
 \cline{3-4}\cline{5-6}\cline{7-8}\cline{9-10} \cline{11-12}
 tag& Eff(\%) & Before & After &   Predict   &Select & Before & After &
                                                                      Predict &   Select& Predict & Select \\
    &         &   NN   &  NN   &            &         &  NN    &  NN  &       &         &         &         \\
\hline
 e         &1.27 &  889.7 &  44.7    &  0.7$\pm$0.2& 0.0 &  889 &  52 & 2.3$\pm$0.4 & 3 &  1.0$\pm$0.2 & 1 \\
 e$\gamma$ &0.18 &  142.4 &   2.7    &  0.0$\pm$0.0& 0.0 &  140 &   6 & 0.2$\pm$0.1 & 0 & 0.1$\pm$0.1 & 0\\
 m         &1.31  & 1486.2 &  59.3   &  1.7$\pm$0.2& 1.1 & 1379 &  63 & 4.3$\pm$0.6 & 2 & 1.9$\pm$0.3 & 1\\
 h         &0.89  &  303.5 &  18.1   &  0.4$\pm$0.1& 0.0 &  318 &  29 & 1.1$\pm$0.3 & 1 &  0.5$\pm$0.1 & 0\\
 h$\gamma$ &2.57 & 1221.9 & 104.3    &  1.7$\pm$0.2& 1.7 & 1160 &  88 & 4.1$\pm$0.6 & 5 &  1.8$\pm$0.3 & 1\\
 3h        &1.22  &  665.4 &  62.2   &  0.6$\pm$0.1& 0.0 &  603 &  74 & 2.1$\pm$0.4 & 1 & 0.9$\pm$0.3 & 1\\
\hline
all tags  & 7.45 & 4709.1 & 291.3    &  5.1$\pm$0.4& 2.8 & 4489 &312  &14.0$\pm$1.0 &12 & 6.2$\pm$0.5& 4 \\
\hline
\end{tabular}}}
\label{tab:nnmubit7-2sig}
\end{center}
\end{table*}
%\end{sidewaystable}

Alternatively, these (and other potential sources of systematic uncertainty not necessarily
accounted for in the above procedure) can be collectively estimated
from the uncertainty in the modelling of the detector by comparing data to the MC backgrounds
in the non-blinded part of GSB, where the background and signal have similar properties
apart from \mec and \DeltaE.
The statistical precision of this data-to-MC ratio is augmented by using 
an expanded \mec range $\in [1.0, 2.5] \gevcc$, where one selects 747 and 713 events
in data and MC backgrounds, respectively. The error on this ratio \datatomcegsb
is ascribed to the systematic error on efficiency, where the 2.3\% normalization 
error arises from the error on the product $\L\sigma_{\tau\tau}$.

The tracking and calorimetry systematic errors affect primarily the uncertainties introduced
 by applying the final signal region requirement. 
 These systematic effects are studied using a control sample of
 \eemm events with energetic photons.
 In order to assess uncertainty on the efficiency arising from the 
 mass and energy scale and resolution systematic errors, 
 the \DeltaE and \mec peaks were varied by $\pm4\mev$ and their resolution by $\pm1\mev$.
 Adding the deviations induced by these variations in quadrature yields an uncertainty of 6.3\% on the efficiency.
 In addition, we estimate the uncertainty on the efficiency associated with the bean energy systematic error
 to be 0.6\%. The total systematic uncertainty on the efficiency is 8.7\%.

 This analysis has an efficiency of \sigeff and an
 expected background, determined from data estimates of the number of background
events in the 2$\sigma$ signal ellispe, $N_{bkg}$, of  \sigback.
After unblinding we find 12 events in the blinded region (to be compared with
14.0$\pm$1.0 expected) and
 \EventsInBox events in the signal region of our data sample.
The probability of 6.2  fluctuating down to \EventsInBox events or fewer
is 26\,\%.
 Table~\ref{tab:nnmubit7-2sig} lists
how these \EventsInBox events are distributed across the different tag channels.
There is good agreement between the observations and expectations in all tag channels
in the unblinded and signal regions.

We calculate the branching fraction of the \taumg decay based on a likelihood function,
which convolutes a Poisson distribution with two Gaussian resolution functions for the background and the efficiency:
\begin{footnotesize}
\begin{equation}
{\cal L}(n, \hat b, \hat f; B, b, f) = {\mu^n e^{-\mu} \over n!}
{1\over 2\pi\sigma_b\sigma_f}e^{-{1\over 2}\left({\hat b - b\over\sigma_b}\right)^2-{1\over2}
\left({\hat f - f\over\sigma_f}\right)^2}
\end{equation}
\end{footnotesize}
where B denotes the branching fraction of (\taumg),
$f=2N_{\tau\tau}\epsilon$, $b$ is the expected total background,
$\mu=\langle n\rangle=fB+b$, $n$ is the number of observed events,
and $\hat b$ ($\hat f$) is sampled from a normal distribution $N(b,\sigma_b)$ ($N(f,\sigma_f)$).
The number of tau pair events  $N_{\tau\tau}$ is \ntaupair.
The errors on the efficiency and normalization are incorporated in $\sigma_f$.
This yields a branching fraction of $\BRtaumg = (-7.5^{+8.1}_{-6.2})\times10^{-8}$.

Since we have no evidence for a signal we have computed an upper limit.
Using the method of Feldman and Cousins\cite{Feldman:1997qc}
the upper limit is 9$\times10^{-8}$ at 90\%CL.
The systematic errors do not change the limit for the number of significant figures quoted.
\footnote{Following a Bayesian approach, the upper limit using a uniform prior in the
branching fraction, the background, and efficiency is \UpperLimitBayes.}

This preliminary result reduces by more than a factor of three the  current upper limit
on the lepton flavour violating decay \taumg
established by the BELLE Collaboration.

%\begin{table}[thb]
%\begin{center}
%\caption{\small {Summary table of the number of expected background events in the signal region,
%signal efficiency and the upper
%limit sensitivity on the decay branching ratio, assuming that the number of observed events
%equals the number
%of expected events. \label{tab:results}}}
%\begin{tabular}{lcc}
%\hline
%\hline
%$N_{\tau\tau}$       & ~~~ & \ntaupair \\
%
%Total background     & ~~~ & \sigback  \\
%
% efficiency        & ~~~ & \sigeff \\
%\hline
%Sensitivity \BRtaumg    & ~~~ & $<$ 19 $\times$ 10$^{-8}$ \\
%\hline
%\BRtaumg                & ~~~ & $<$ 0.9 $\times$ 10$^{-8}$ \\
%\hline
%\hline
%\end{tabular}
%\end{center}
%\end{table}

% Input the pubboard acknowledgements file
%\noindent {\bf ACKNOWLEDGEMENTS}\\ 
%\input acknow_PRL.tex


\begin{thebibliography}{99}
\bibitem{flatphasespace}
Although the signal MC has been modelled using a flat phase space model, the limit obtained
in this analysis is insensitive to this assumption as demonstrated by considering the two 
extreme cases of a V-A and a V+A form of interaction for the signal MC.
\end{thebibliography}
\end{document}